\documentclass{article}
\usepackage{graphics,latexsym}
\usepackage{graphicx}
\usepackage{amsmath, amssymb,amsthm}

\begin{document}

\title{Electromagnetic Radiations \\ as a Fluid Flow}

\author{Daniele Funaro \\ Department of Mathematics
\\ University of Modena and Reggio Emilia \\ Via Campi 213/B, 41125 Modena (Italy)
\\ daniele.funaro@unimore.it}

\date{}
\maketitle

\begin{abstract}
We combine Maxwell's equations with Eulers's equation, related to
a velocity field of an immaterial fluid, where the density of mass
is replaced by a charge density. We come out with a differential
system able to describe a relevant quantity of electromagnetic
phenomena, ranging from classical dipole waves to solitary
wave-packets with compact support. The clue is the construction of
an energy tensor summing up both the electromagnetic stress and a
suitable mass tensor. With this right-hand side, explicit
solutions of the full Einstein's equation are computed for a wide
class of wave phenomena. Since our electromagnetic waves may
behave and interact exactly as a material fluid, they can create
vortex structures. We then explicitly analyze some vortex ring
configurations and examine the possibility to build a model for
the electron.
\end{abstract}

\vspace{.9cm}



%47.32.cf (vortex rings in fluids)
%12.10.-g (Unified field theories and models)
% 03.50.De (Classical electromagnetism, Maxwell equations)
% 04.20.Jb (general relativity, exact solutions)
\renewcommand{\theequation}{\arabic{equation}}

\medskip
\section{Introduction}
Since the advent of the theory of electromagnetic fields, more
than a century ago, waves have been described as a kind of energy
flow, governed by suitable transport equations in vector form,
namely Maxwell's equations. The original publications (see
\cite{maxwell1} and \cite{maxwell2}), in a way that nowadays may
seems naif, actually refer to electric and magnetic phenomena
using mechanical terms, such as tension, stress, pressure and
vortices, in association with some `imaginary substance'.
\par\smallskip

It turns out that, in void, the electric and magnetic fields
(${\bf E}$ and ${\bf B}$, respectively) are transversally oriented
with respect to the direction of propagation, and their envelope
produces a sequence of wave-fronts. This is in agreement with  the
fact that the energy develops according to the evolution of the
vector product ${\bf E}\times {\bf B}$, otherwise known as
Poynting vector.
\par\smallskip

On the other hand, the dynamical behavior of a compressible non
viscous fluid is well described by Euler's equation, where, in
principle, the velocity vector field (denoted by ${\bf V}$) might
not be necessarily related to a real material fluid. In
particular, one could replace the mass density by a sort of charge
density. Therefore, the temptation to describe electromagnetic and
velocity fields, through a combination of the respective modelling
equations, is well motivated.
\par\smallskip

We are going to present a system of equations in the three
independent vector unknowns: $({\bf E}, {\bf B}, {\bf V})$. In
pure void, the electric and magnetic fields follow the Faraday's
law together with the Amp\`ere's law, where a current, flowing at
velocity ${\bf V}$, is supposed to be naturally associated with
the wave. In order to close the system, the third relation is the
Euler's equation for ${\bf V}$, containing an added forcing term
${\bf E}+{\bf V}\times {\bf B}$, perfectly analogous to that
characterizing the Lorentz's law. In this way, the three entities
$({\bf E}, {\bf B}, {\bf V})$ turn out to be strictly entangled.
\par\smallskip

Despite the appearance, the new model allows for a very large
space of solutions. Moreover, it displays numerous conservation
and invariance properties, all deducible from a standard analysis.
An interesting invariant subspace of solutions is the one where
the third equation is reduced to ${\bf E}+{\bf V}\times {\bf
B}=0$, which means that no acceleration is acting on the wave, and
the corresponding `flow' is somehow laminar. For this
circumstance, the solutions, called free-waves, perfectly follow
the laws of geometrical optics, ruled by the eikonal equation.
Together with other known solutions, free-waves also include
solitary electromagnetic waves with compact support almost of any
shape, intensity, frequency and polarization. Such a result is
clearly important, since it reopens the path to a serious
discussion, based on a deterministic analysis, on crucial issues
as the structure of photons, the duality wave-particle and the
quantum properties of matter.
\par\smallskip

Far more complicated solutions (not of the free-wave type) are
however possible. Since our electromagnetic radiations actually
behave as a fluid, they can be constrained to evolve in bounded
regions of space, similar for instance to vortex rings. According
to the model equations, rotating photons in a vortex structure may
carry a charge and deform, via Einstein's equation, the local
geometry of the space-time in order to create a gravitational
environment assimilable to the presence of mass. The same metric
space is responsible for the stability of such a wave, obliged in
this way to develop along self-created geodesics. This leaves us
with the conjecture that some stable elementary particles (as for
instance the electron) could be made by rotating photons, an idea
that has been put forward by many authors in the past, although
with not too much recognition, basically due to the lack of a
sufficient theoretical description of electromagnetic phenomena,
able to go beyond the classical linear Maxwellian approach. With
the help of the new material collected here, we examine this
aspect in the last section.
\par\smallskip

This paper is a review and a reorganization of the results
presented in \cite{funarol}, and contains additional new material.
The presentation follows an inverse path, and the equations are
first derived in covariant form and successively deduced in the
flat space. The general relativity framework is however the key
point to understand the structure of the solutions, especially in
the case of constrained waves. The entire theory comes out as a
consequence of a wise choice of the right-hand side tensor in the
formulation of Einstein's equation. This combines in the proper
manner the electromagnetic stress tensor and a mass type tensor,
where the density is referred to the electric field. In the
subspace of free-waves, by generalizing the results given in
\cite{funarol}, we are able to write down explicit general
solutions of the Einstein's equation. As a final step, we will
derive a notion of mass density, that is going to be zero in the
average for free-waves (and photons) and positive for electrons.

\par\smallskip

\section{The right-hand side tensor}

Let us fix the notation by introducing the Einstein's equation:
\begin{equation}\label{eq:eins}
R_{\alpha\beta}~-~{\textstyle\frac{1}{2}}g_{\alpha\beta}R~=~ -\chi
T_{\alpha\beta}
\end{equation}
where $R_{\alpha\beta}$ is the Ricci tensor, $R$ the scalar
curvature (the trace of $R_{\alpha\beta}$) and $\chi$ a positive
adimensional constant. The signature of the metric space
$g_{\alpha\beta}$ will be $(+,-,-,-)$. The given constant $\chi$,
that will be estimated at the end of section 6, has nothing in
common with the gravitational constant $G$.
\par\smallskip

As anticipated in the introduction, the main ingredient of
our theory, is the choice of the right-hand side tensor. We set:
\begin{equation}\label{eq:tens}
T_{\alpha\beta}~=~\frac{\mu}{c^4}\Big( -\mu U_{\alpha\beta}
+M_{\alpha\beta}\Big)
\end{equation}
where $c$ is the speed of light and $\mu$ is a constant,
dimensionally equivalent to a charge divided by a mass
(see (\ref{eq:vum})). In
(\ref{eq:tens}) we find the usual electromagnetic stress tensor:
\begin{equation}\label{eq:elec}
U_{\alpha\beta}~=~ -g^{\gamma\delta}F_{\alpha\gamma}F_{\beta\delta}
+{\textstyle{\frac{1}{4}}}g_{\alpha\beta}F_{\gamma\delta}F^{\gamma\delta}
\end{equation}
with
\begin{equation}\label{eq:etens}
F_{\alpha\beta}~=~ \nabla_\alpha A_\beta -\nabla_\beta A_\alpha
\end{equation}
where $A_\alpha$ is the electromagnetic  4-vector potential
(energy/charge) and $\nabla_\alpha$ denotes the covariant derivative.
Although it is not strictly requested, $A_\alpha$
satisfies the Lorenz gauge relation:
\begin{equation}\label{eq:lorenz}
\nabla_\alpha A^\alpha~=~0
\end{equation}
In (\ref{eq:tens}) we also find a kind of mass tensor:
\begin{equation}\label{eq:mass}
M_{\alpha\beta}~=~ \rho V_\alpha V_\beta - g_{\alpha\beta}p
\end{equation}
where $V_\alpha$ is velocity 4-vector, $\rho$ plays  the role of a
density  and $p$ is similar to a pressure. Dimensionally, $\rho$
and $p$ are taken in order to conform with the setting in
(\ref{eq:tens}), therefore they will not correspond to the usual
entities of classical fluid dynamics. Moreover, differently from
what it is generally  supposed (see for instance \cite{fock},
p.91), no direct relations between $\rho$ and $p$ are a priori
requested.
\par\smallskip

This way of building $T_{\alpha\beta}$ characterizes the whole
theory. We remark (and this is an important key point) that
$U_{\alpha\beta}$ in (\ref{eq:tens}) appears with the opposite
sign. The reason for this choice will be clear as we proceed in
the exposition.
\par\smallskip

Concerning the vector field $V_\alpha$, we shall make the following
assumption:
\begin{equation}\label{eq:eiko}
g^{\alpha\beta}V_\alpha V_\beta~=~0
\end{equation}
that is equivalent to an eikonal equation for suitable propagating
wave-fronts.
\par\smallskip

A first important relation is obtained by evaluating the
trace of (\ref{eq:eins}). This gives:
\begin{equation}\label{eq:press}
p~=~\frac{c^4}{4\chi\mu}R
\end{equation}
In computing (\ref{eq:press}) we considered  that both the traces
of $U_{\alpha\beta}$ and $\rho V_\alpha V_\beta$ are zero. In the
first case this is a known property of the electromagnetic stress
tensor. In the second case it is a straightforward consequence of
relation (\ref{eq:eiko}). Equation  (\ref{eq:press}) relates
curvature and pressure. Wave-fronts travelling unperturbed,
without developing interstitial pressure, actually move in a space
with zero scalar curvature, and vice versa (see section 4).

\par\smallskip
Let us then derive from (\ref{eq:eins}) a set of covariant
equations. In order to have compatibility with the  left-hand side
in (\ref{eq:eins}), the most important relation to be satisfied
is:
\begin{equation}\label{eq:divt}
\nabla_\beta T^{\alpha\beta}~=~0
\end{equation}
yielding the set of modelling equations. Due to the special
expression of $T_{\alpha\beta}$, the energy conservation property
(\ref{eq:divt}) mixes up electromagnetic and mechanical terms in a
very natural way. Thus, we start by observing that (see for
instance \cite{fock} or \cite{misner}):
\begin{equation}\label{eq:divum0}
\nabla_\beta U^{\alpha\beta}~=~ g^{\alpha\gamma}F_{\gamma\delta}(\nabla_\beta
F^{\delta\beta})
\end{equation}
\begin{equation}\label{eq:divum}
\nabla_\beta M^{\alpha\beta}~=~\rho G^\alpha + \nabla_\beta (\rho
V^\beta ) V^\alpha - g^{\alpha\beta}\nabla_\beta p ~~~~{\rm
with}~~ G^\alpha = V^\gamma \nabla_\gamma V^\alpha
\end{equation}
Thus, according to (\ref{eq:divt}), by summing up the above expressions,
one must have:
\begin{equation}\label{eq:divtc}
-~\mu g^{\alpha\gamma}F_{\gamma\delta}(\nabla_\beta
F^{\delta\beta})~+~\rho G^\alpha~+~\nabla_\beta (\rho V^\beta )
V^\alpha ~-~g^{\alpha\beta}\nabla_\beta p~=~0
\end{equation}
By adding and subtracting the  term $~(\mu\rho
/c)g^{\alpha\gamma}F_{\gamma \delta}V^\delta =(\mu\rho /c)
F^{\alpha\delta}V_\delta$, equation (\ref{eq:divtc}) becomes:
$$\mu g^{\alpha\gamma}F_{\gamma\delta}\Big(-\nabla_\beta
F^{\delta\beta}-\frac{\rho}{c} V^\delta \Big)~+~\rho \Big(G^\alpha
+\frac{\mu}{c}F^{\alpha\delta}V_\delta \Big)~~~~~~~~~~~~~~$$
\begin{equation}\label{eq:divtcm}
~~~~~~~~~~~~~~~~+~\nabla_\beta (\rho V^\beta ) V^\alpha
~-~g^{\alpha\beta}\nabla_\beta p~=~0
\end{equation}
\par\smallskip

Hence, in order to obtain (\ref{eq:divtcm}), a  sufficient
condition is to impose the following set of equations:
\begin{equation}\label{eq:maxm}
c ~\nabla_\beta F^{\delta\beta}~=~-\rho V^\delta
\end{equation}
\begin{equation}\label{eq:fluid}
\rho \Big(G^\alpha  +\frac{\mu}{c}F^{\alpha\delta}V_\delta \Big)~
=~\nabla^\alpha p
\end{equation}
that, being written in covariant form, are, in principle,
admissible in any geometric environment, but, hopefully, in the
metric spaces compatible with relation (\ref{eq:eiko}). Equation
(\ref{eq:maxm}) is the Amp\`ere law and (\ref{eq:fluid}) is the
Euler's equation for a certain velocity field (not corresponding
to a material fluid). In truth, in order to get (\ref{eq:divtcm}),
one should also impose the following continuity  equation:
\begin{equation}\label{eq:conti}
\nabla_\beta (\rho V^\beta )~=~0
\end{equation}
but this is a consequence of equation (\ref{eq:maxm}). In fact, by
taking the 4-divergence of (\ref{eq:maxm}),  one has:
$c\nabla_\delta(\nabla_\beta F^{\delta\beta})=0=\nabla_\delta(\rho
V^\delta )$, resulting from the anti-symmetry of the tensor
$F^{\delta\beta}$.
\par\smallskip

There is an interesting compatibility relation to be remarked.
First, thanks to (\ref{eq:eiko}), one recovers that:
\begin{equation}\label{eq:ortog}
G^\alpha V_\alpha ~=~{\textstyle{\frac{1}{2}}}(G^\alpha V_\alpha+
G_\alpha V^\alpha
)~=~{\textstyle{\frac{1}{2}}}(V^\gamma\nabla_\gamma )(V_\alpha
V^\alpha )~=~0
\end{equation}
Then, due to the anti-symmetry of $F^{\delta\beta}$,
(\ref{eq:fluid}) implies:
\begin{equation}\label{eq:presco}
 V_\alpha \nabla^\alpha p~=~0
\end{equation}
which is a sort of conservation property for $p$ (a kind of
Bernoulli's principle, if one considers that $p=p+\frac{1}{2}\rho
V_\alpha V^\alpha$).

\par\smallskip
The equations (\ref{eq:maxm}) and (\ref{eq:fluid}), together with
the {\sl equation of state} (\ref{eq:press}), are the foundations
of our theory. In principle, one could obtain (\ref{eq:divt})
without necessarily enforcing (\ref{eq:maxm}) and
(\ref{eq:fluid}). However, we believe that all the most relevant
phenomena can be described through the two differential equations
written above.
\par\smallskip

\section{The model equations in Minkowski space}

To better understand what  is happening we can specialize
equations (\ref{eq:maxm}) and (\ref{eq:fluid}) in the case of a
flat metric space $~g_{\alpha \beta}={\rm diag}\{ 1,-1,-1,-1\}$,
in the Cartesian coordinate system $(ct,x_1,x_2,x_3)$.
\par\smallskip

We start from the potential $(A_0,A_1,A_2, A_3)=(-\Phi, {\bf A})$
and through (\ref{eq:etens}), considering that $\nabla_\alpha =(
\frac{1}{c}\frac{\partial}{\partial t}, \nabla )$, we obtain the
electromagnetic tensor (see \cite{fock}, section 24, or
\cite{atwater}, section 5.2):
\begin{equation}\label{eq:tens2}
F_{\alpha\beta}~=~\left(\begin{array}{c @{\hspace{.5cm}} c
@{\hspace{.5cm}} c @{\hspace{.5cm}} c} 0 & -E_1 & -E_2 & -E_3 \\
E_1 & 0 & cB_3 & -cB_2 \\ E_2 & -cB_3 & 0 & cB_1 \\ E_3 & cB_2 &
-cB_1 & 0 \end{array}\right)
\end{equation}
where ${\bf E}=(E_1,E_2,E_3)$ is the standard electric field and
${\bf B}=(B_1,B_2,B_3)$ is the magnetic field. The above
construction is true by virtue of the setting (see
(\ref{eq:etens})):
\begin{equation}\label{eq:ebpote}
{\bf B}~=~\frac{1}{c}~{\rm curl}{\bf A}~~~~~~~~~~~~~
{\bf E}~=~-\frac{1}{c}\frac{\partial \bf A}{\partial t}~-~\nabla\Phi
\end{equation}
\par\smallskip

\noindent In particular, we have:
\begin{equation}\label{eq:sfbm1}
\frac{\partial {\bf B}}{\partial t}~=~ -{\rm curl} {\bf E}~~~~~~~~~~~~~~
~~~~{\rm div}{\bf B} ~=~0
\end{equation}
as a direct consequence of (\ref{eq:ebpote}). The contravariant
tensor $F^{\alpha\beta}$ is obtained from $F_{\alpha\beta}$ by
replacing ${\bf E}$ by $-{\bf E}$.
\par\smallskip

Without loss of generality,  we can set $V^0=-c$, so that, from
(\ref{eq:maxm}) with $\delta=0$, we deduce: $\rho =\nabla_\beta
F^{0\beta}={\rm div}{\bf E}$.
Denoting  by ${\bf V}=(V_1,V_2,V_3)$ the velocity field, by
(\ref{eq:eiko}) one obtains $~V_\alpha =(-c, {\bf V})~$ and
$~V^\alpha =(-c, -{\bf V})$. Finally we get:
\begin{equation}\label{eq:sfem1}
\frac{\partial {\bf E}}{\partial t}~=~ c^2{\rm curl} {\bf B}~
-~\rho {\bf V}
\end{equation}
\begin{equation}\label{eq:slor1}
\rho\left(\frac{D{\bf V}}{Dt}~+~\mu ({\bf E}+{\bf V}\times {\bf B})
\right) ~=~-\nabla p
\end{equation}
\begin{equation}\label{eq:epres}
\frac{\partial p}{\partial t}~=~\mu\rho ({\bf E}\cdot {\bf V})
\end{equation}
where $~\frac{D}{Dt}{\bf V}=\frac{\partial}{\partial t}{\bf
V}+({\bf V}\cdot \nabla ){\bf V}~$  is the substantial derivative,
so that (\ref{eq:slor1}) ends up to be the Euler's equation.
Here equation (\ref{eq:sfem1}) comes from (\ref{eq:maxm})  for
$\delta =1,2,3,$  equation (\ref{eq:slor1}) comes from
(\ref{eq:fluid}) for $\alpha =1,2,3,$ and equation
(\ref{eq:epres}) comes from (\ref{eq:fluid}) for $\alpha =0$. In
addition, one has the continuity equation:
\begin{equation}\label{eq:continu}
\frac{\partial \rho}{\partial t}~=~-{\rm div}(\rho {\bf V})
\end{equation}
that is easily derived either from (\ref{eq:conti}) or by taking
the divergence of (\ref{eq:sfem1}).
\par\smallskip

The equations (\ref{eq:press}) and (\ref{eq:eiko}) are not valid
in  this context, since they are strongly related to the fact that
$g_{\alpha\beta}$ is solution of (\ref{eq:eins}), which is not
true in the simplified case we are examining. The solutions to
(\ref{eq:sfem1})-(\ref{eq:epres}) are however important to set up
the tensor $F_{\alpha\beta}$ in (\ref{eq:etens}), which may be
computed using partial derivatives in place of covariant
derivatives. This enables us to build the tensor (\ref{eq:tens})
for a generic $g_{\alpha\beta}$ to be used as a right-hand side in
(\ref{eq:eins}).
\par\smallskip

An even more simplified version is obtained by requiring
$D{\bf V}/Dt=0$ and $p=0$ in (\ref{eq:slor1}):
\begin{equation}\label{eq:sfem2}
\frac{\partial {\bf E}}{\partial t}~=~ c^2{\rm curl} {\bf B}~
-~\rho {\bf V}
\end{equation}
\begin{equation}\label{eq:sfbm2}
\frac{\partial {\bf B}}{\partial t}~=~ -{\rm curl} {\bf E}
\end{equation}
\begin{equation}\label{eq:sfdb2}
{\rm div}{\bf B} ~=~0
\end{equation}
\begin{equation}\label{eq:slor2}
{\bf E}~+~{\bf V}\times {\bf B} ~=~0
\end{equation}
where $\rho ={\rm div}{\bf E}$, and ${\bf V}$ is a velocity vector
field satisfying $\vert{\bf V}\vert =c$. Note that (\ref{eq:eiko})
is now true. The field ${\bf V}$ is oriented as the vector field
${\bf E}\times {\bf B}$. Now (\ref{eq:press}) is also correct,
since the property $R=0$ is compatible with the flatness of the
space. Note that relation (\ref{eq:slor2}) is certainly satisfied
for all electromagnetic waves where ${\bf E}$ is orthogonal to
${\bf B}$ and $\vert {\bf E}\vert =\vert c{\bf B}\vert$. Moreover,
from (\ref{eq:slor2}) we get ${\bf E}\cdot {\bf V}=0$, that is in
agreement with (\ref{eq:epres}), since $p$ is identically zero.
Take into account, however, that the flat metric does not satisfy
(\ref{eq:eins}), since in this case one has $R_{\alpha\beta}=0$,
in contrast to the fact that $T_{\alpha\beta}\not =0$.
\par\smallskip

This set of equations produces a  closed subset of solutions
called {\sl free-waves}, that contains all the electromagnetic
phenomena in vacuum travelling undisturbed according to the rules
of geometrical optics. Indeed, if ${\bf V}=\nabla\Psi$ is a
gradient, then the relation $\vert{\bf V}\vert =\vert
\nabla\Psi\vert =c$ is the eikonal equation. This subspace
includes solitary waves with compact support, perfect spherical
waves, and many other solution non obtainable with the classical
Maxwell's setting. In this case, equation (\ref{eq:sfem2}) comes
from finding the stationary points of the standard Lagrangian of
the electromagnetism, under the condition $~A_\alpha V^\alpha =0~$
on the potentials (see \cite{funarop}). Moreover, (\ref{eq:sfem2})
is invariant under Lorentz transformations (see \cite{funarol},
section 2.6).
\par\smallskip

The subset of free-waves is characterized in covariant form by the
following expression, deduced from (\ref{eq:fluid}) for $p=0$ and
$G^\alpha =0$:
\begin{equation}\label{eq:fluidf}
F^{\alpha\delta}V_\delta ~=~0
\end{equation}
that brings to (\ref{eq:slor2}) and ${\bf E}\cdot {\bf V}=0$. It
can also be proven that both (\ref{eq:continu}) and
(\ref{eq:fluidf}) are invariant under Lorentz transformations.
\par\smallskip

The triplet $({\bf E},{\bf B},{\bf V})$  turns out to be
right-handed (in the right-handed reference frame $(x,y,z)$).
There is however an alternative path, that brings to left-handed
triplets. It is enough to replace (\ref{eq:ebpote}) by:
\begin{equation}\label{eq:ebpotea}
{\bf B}~=~-\frac{1}{c}~{\rm curl}{\bf A}~~~~~~~~~~~~~ {\bf
E}~=~-\frac{1}{c}\frac{\partial \bf A}{\partial t}~-~\nabla\Phi
\end{equation}
and rewrite all the equations with $-{\bf B}$ in place of ${\bf
B}$. We get in this way a completely new set of electromagnetic waves
that have a specular image with respect to the classical one.
\par\smallskip

\section{Explicit solutions}

In the case of free-waves explicit expressions of the metric
tensor $g_{\alpha\beta}$ are available. The results we are going
to show generalize  those given in \cite{funarol}.
\par\smallskip

Let us assume the following form for the potentials:
\begin{equation}\label{eq:potes}
A_\alpha ~=~\Big(-f(x,y)\tau (\xi ),~0,~0,~f(x,y)\tau (\xi )\Big)
\end{equation}
Here we are in Cartesian coordinates
$(x_0,x_1,x_2,x_3)=(ct,x,y,z)$ and we set $\xi = z-ct$. The two
functions $f$ and $\tau$ are arbitrary.
\par\smallskip

With this setting, we obtain the electromagnetic fields:
\begin{equation}\label{eq:fies}
{\bf E}=(-f_x\tau , ~-f_y\tau ,~0)~~~~~~~~~~~~{\bf ~B}=c^{-1}(f_y\tau ,
~-f_x\tau ,~0)
\end{equation}
where $f_x=\partial f/\partial x$ and $f_y=\partial f/\partial y$.
Thus, we are in presence of fronts parallel to the plane $(x,y)$
and evolving in the direction of the $z$ axis at speed $c$. Plane
waves are obtained when $f$ is of the form $ax+by$, for some
constants $a$ and $b$. The case when $f$ has compact support is
quite interesting, since it give rise to self-contained solitary
waves, travelling straightly at the speed of light (photons).
These solutions do not belong to the Maxwellian theory (see also
\cite{funarop2} and \cite{funarop}). By defining $V_\alpha =(V_0,
{\bf V})=(-c,0,0,c)$, the entire set of equations
(\ref{eq:sfem2})-(\ref{eq:slor2}) is satisfied in the flat space.
Thus, we can also write $\xi=z-V_3t$. We observe that ${\rm
div}{\bf B}$ is automatically zero, while we do not expect $~{\rm
div}{\bf E}=-(f_{xx}+f_{yy})\tau~$ to be zero.
\par\smallskip

For any arbitrary (nonzero) function $s$ of the variable $x+y$, we
look for a metric tensor having the following structure:
$$g_{00}=1~~~~~~~g_{11}=-f^2_x\sigma^2-s^2~~~~~~g_{12}=g_{21}=-f_xf_y\sigma^2-s^2$$
\begin{equation}\label{eq:gtens}
~g_{22}=-f^2_y\sigma^2-s^2~~~~~~~~g_{33}=-1
\end{equation}
being the other entries equal to zero. Here $\sigma$ is a function
of the variable $\xi$. Note that, in this situation and in accordance
to (\ref{eq:eiko}), we  have:
\begin{equation}\label{eq:vaz}
V^\alpha V_\alpha ~=~0~~~~{\rm and}~~~~V^\alpha A_\alpha ~=~0
\end{equation}
with $V^\alpha =(-c,0,0,-c)$.
\par\smallskip

The determinant of the metric tensor is: $g=-[s(f_x-f_y)\sigma
]^2$. For this reason, it is necessary that $s\not =0$. We have:
\begin{equation}\label{eq:tensf}
F_{\alpha\beta}=\tau \left(\begin{array}{c @{\hspace{.2cm}} c
@{\hspace{.2cm}} c @{\hspace{.2cm}} c} 0 & f_x & f_y & 0 \\
-f_x & 0 & 0 & f_x \\ -f_y & 0 & 0 & f_y \\ 0 & -f_x &
-f_y & 0 \end{array}\right)~~~~~~
F^{\alpha\beta}=\frac{\cal F}{\sqrt{-g}}
\left(\begin{array}{c @{\hspace{.2cm}} c
@{\hspace{.2cm}} c @{\hspace{.2cm}} c} 0 & -1 & +1 & 0 \\
+1 & 0 & 0 & +1 \\ -1 & 0 & 0 & -1 \\ 0 & -1 &
+1 & 0 \end{array}\right)
\end{equation}
where ${\cal F}=\pm \tau s/\sigma$, and the sign is set up
according to that of $s(f_x-f_y)\sigma$.
\par\smallskip

By explicit computation, one can get the Ricci tensor, where the
nonzero entries are:
\begin{equation}\label{eq:riccic}
R_{00}=R_{33}=-R_{30}=-R_{03}=-\sigma^{\prime\prime}/\sigma
\end{equation}
In addition, one gets $R=0$. Amazingly, the functions $s$ and $f$
disappear. For completeness, we also report the nonzero
Christoffel symbols, for $s^2$ constantly equal to 1:
$$\Gamma^0_{11}=\Gamma^3_{11}=-f^2_x \sigma\sigma^\prime~~~~~~~~~
\Gamma^0_{22}=\Gamma^3_{22}=-f^2_y \sigma\sigma^\prime$$
$$\Gamma^0_{12}=\Gamma^0_{21}=\Gamma^3_{12}=\Gamma^3_{21}=-f_xf_y \sigma\sigma^\prime$$
$$\Gamma^1_{11}=-\Gamma^2_{11}=f_{xx}/(f_x-f_y)~~~~~~~
\Gamma^1_{22}=-\Gamma^2_{22}=f_{yy}/(f_x-f_y)$$
$$\Gamma^1_{12}=\Gamma^1_{21}=-\Gamma^2_{12}=-\Gamma^2_{21}=f_{xy}/(f_x-f_y)$$
$$-\Gamma^1_{01}=-\Gamma^1_{10}=\Gamma^1_{13}=\Gamma^1_{31}=
-\Gamma^2_{31}=-\Gamma^2_{13}=\Gamma^2_{01}=\Gamma^2_{10}=\frac{f_x\sigma^\prime}
{\sigma (f_x-f_y)}$$
\begin{equation}\label{eq:chris}
-\Gamma^1_{02}=-\Gamma^1_{20}=\Gamma^1_{23}=\Gamma^1_{32}=
-\Gamma^2_{32}=-\Gamma^2_{23}=\Gamma^2_{02}=\Gamma^2_{20}=\frac{f_y\sigma^\prime}
{\sigma (f_x-f_y)}
\end{equation}
We should avoid the points $(x,y)$ such that $f_x=f_y$, but this
is in general a set of measure equal to zero. The coefficients of
the metric tensor remain however smooth everywhere, thus the fact
that the determinant $g$ can be zero somewhere is not a crucial
problem.
\par\smallskip

On the other hand, based on the proposed metric, we  can evaluate
the electromagnetic stress tensor (\ref{eq:elec}), where the
nonzero entries are:
\begin{equation}\label{eq:utens}
U_{00}=U_{33}=-U_{30}=-U_{03}=\tau^2/\sigma^2
\end{equation}
In the new metric space, one also discovers that:
\begin{equation}\label{eq:caro}
\rho =\nabla_\beta F^{0\beta}=\frac{1}{\sqrt{-g}}\frac{\partial}
{\partial x_\beta}(\sqrt{-g}~ F^{0\beta })=\frac{1}{\sqrt{-g}}
\left(-\frac{\partial {\cal F}}{\partial x}+
\frac{\partial {\cal F}}{\partial y}\right)=0
\end{equation}
where we used that $s$ is a function of $x+y$ and $\tau /\sigma$
does not depend on $x$ and $y$. Thus, we can get rid of the mass
tensor in (\ref{eq:tens}). We can also check that
(\ref{eq:fluidf}) is true, ensuring that the concept of free-wave
is, in some sense, invariant. As a further confirmation, one can
also check that (see (\ref{eq:divum})):
\begin{equation}\label{eq:cgz}
G^\alpha ~=~ V^\beta \frac{\partial V^\alpha}{\partial x_\beta}
~+~\Gamma^\alpha_{\gamma\delta}V^\gamma V^\delta~=~0
\end{equation}
because $V^\alpha =(-c,0,0, -c)$ is constant and the Christoffel
symbols $\Gamma^\alpha_{00}$, $\Gamma^\alpha_{03}$,
$\Gamma^\alpha_{30}$ and $\Gamma^\alpha_{33}$ are zero. The above
properties are compatible with the choice $p=0$.
\par\smallskip

In the end, putting together (\ref{eq:riccic}) and
(\ref{eq:utens}), we find out that $g_{\alpha \beta}$ is solution
of the Einstein's equation (\ref{eq:eins}), if the following
differential equation in the variable $\xi$ is verified:
\begin{equation}\label{eq:eqxi}
-\sigma \sigma^{\prime\prime}~=~\frac{\chi\mu^2}{c^4}\tau^2
\end{equation}
For example, when $\tau (\xi )=\sin\omega\xi$, equation
(\ref{eq:eqxi}) admits the simple solution given by:
\begin{equation}\label{eq:solsig}
\sigma (\xi )~=~\frac{\sqrt{\chi}\mu}{c^2\omega}\sin\omega\xi
\end{equation}
Therefore, $g_{\alpha \beta}$ is associated with a monochromatic
gravitational wave strictly tight to the given electromagnetic
wave, and having the same frequency and support. The intensity of
$\sigma$ is inversely proportional to the frequency. This somehow
agrees with the observation that gravitational phenomena are more
relevant when one is dealing with very low frequencies. If we
could ride a wave, we would not be able to `see' in the
transversal direction, since, in the modified metric, the tensors
$R_{\alpha\beta}$ and $U_{\alpha\beta}$ (see (\ref{eq:riccic}) and
(\ref{eq:utens}), respectively) do not depend on $x$ and $y$. This
justifies why in this case we have $\rho =0$ (independently of the
value ${\rm div}{\bf E}$ originally attributed in the flat space).
This means that, in its geometry, a photon looks like an unbounded
plane wave. Note that here we solved the Einstein's equation in
full form, and not its linearized version, as it is usually done
in other contexts. As pointed out in \cite{aldrovandi} this is a
decisive and necessary improvement.
\par\smallskip

By coordinate transformations, other types of wave-fronts (for
instance of spherical shape) can be studied in a very similar
fashion. Outside the support of a wave, there is no signal. There
the metric space is flat, compatibly with the metric tensor
(\ref{eq:gtens}) when one sets $f_x=0$ and $f_y=0$. Basically, for
all the family of free-waves, we can explicitly provide a full
solution to (\ref{eq:eins}), including situations (not
investigated here, but easy to handle) where the polarization
varies with $\xi$. Such a general result extends the analysis
developed in \cite{funarol}. For $f_x$ and $f_y$ constant we get
(unbounded) plane waves. Gravitational plane waves in vacuum were
firstly found in \cite{bondi} (see also \cite{misner}, section
35). The results are related to a situation different from the one
presented here, and the waves look actually flat after the
introduction of a suitable `planeness' concept.
\par\smallskip

We would like to remark that our analysis has been made possible,
in such a clean and elegant way, because the electromagnetic
tensor $U_{\alpha\beta}$ in (\ref{eq:tens}) appears with the
opposite sign.  Our choice can be justified by several reasons.
First, as we said, with the sign adopted here we can come out with
a multitude of interesting and significant exact solutions.
Actually, by changing the sign of the right-hand side in
(\ref{eq:eqxi}), there are a few chances to get reasonable bounded
solutions, without introducing some passages not well justified
from the mathematical point of view. Secondly, when studying more
complex situations where the mass tensor in (\ref{eq:tens}) plays
an effective role, we can interpret Einstein's equation as a
balance law, between electromagnetic and mechanical energies
operating in opposition (there might be some analogies with the
results in \cite{lo}). This is actually what happens for the
electromagnetic vortices we are going to study in the next
sections.

\par\smallskip

Old well-known solutions can be also adapted to the new context.
This is the case for example of the  Reissner-Nordstr\"om metric
(see \cite{misner}, section 33.2). Relatively to the spherical
reference framework $(r,\theta ,\phi )$, we start from the
potential $A_\alpha =(q/r,0,0,0)$. Successively, the nonzero
entries of the metric are defined as:
\begin{equation}\label{eq:rnme}
g_{00}=1-\frac{M}{r}-\frac{Q^2}{r^2}~~~~~g_{11}=-1/g_{00}~~~~g_{22}=-r^2
~~~~ g_{33}=-(r\sin\theta )^2
\end{equation}
for some constants $M>0$ and $Q$ related to the mass and the
charge of the black-hole. The difference with the standard case is
a switch of a sign in the expression of $g_{00}$ (usually this is
equal to $1-M/r+Q^2/r^2$).  One checks that $\rho =0$ and $R=0$
(with the exception of the singular point $r=0$). Moreover, by
taking $Q=q\sqrt{\chi\mu^2/2c^4}$, the new metric in
(\ref{eq:rnme}) turns out to be solution of the Einstein's
equation:
\begin{equation}\label{eq:einrn}
R_{\alpha\beta}~=~\frac{\chi\mu^2}{c^4}U_{\alpha\beta}
\end{equation}
that corresponds to (\ref{eq:eins}) with the right-hand side
tensor (\ref{eq:tens}), in contrast to the classical approach
where $R_{\alpha\beta}=-(\chi\mu^2 /c^4)U_{\alpha\beta}$ (see for
instance \cite{atwater}, p. 117).
\par\smallskip

It is important to remark that, with such a new setting, one can
remove the constraint $M>2Q$, granting in this way the existence
of the horizons at any regime, even with small masses. A similar
modification of the coefficients (that is: $q^2\rightarrow -q^2$)
also works in the case of the Kerr-Newman metric (see
\cite{funarol}).  This also shows that switching  the sign of
$U_{\alpha\beta}$ is not a traumatic choice, but can give rise to
meaningful aspects never investigated before. The possibility of
dealing with electric objects with small masses reopens the path
to the study of gravitational electron models, without the use of
involved geometrical extensions (see \cite{arcos}). We examine
this possibility in section 6.
\par\smallskip

We finally provide a definition of mass density $\rho_m$, which is
derived from the coefficient $M_{00}$ of the mass tensor in
(\ref{eq:mass}). In the case when $g_{\alpha\beta}=\{1,-1,-1,-1\}$
and $\vert V_0\vert =c$, we set:
\begin{equation}\label{eq:made}
\rho_m~=~\frac{\epsilon_0 M_{00}}{\mu c^2}~=~\frac{\epsilon_0}{\mu}\left(
\rho -\frac {p}{c^2}\right)
\end{equation}
If $\Sigma$ is the support of the wave, the corresponding mass $m$
is obtained by integrating $\rho_m$ over $\Sigma$ (we have
$\int_\Sigma \rho_m\sqrt{-g}~$ for a generic metric space). In the
case of a free-wave we have $p=0$. If, in addition the wave has
compact support $\Sigma$, we have: $~m=\int_\Sigma \rho_m
=(\epsilon_0 /\mu)\int_\Sigma \rho =(\epsilon_0 /\mu)
\int_{\partial\Sigma}{\bf E}\cdot {\bf n}=0$. In agreement with
the Gauss's theorem, we are not in presence of classical charges.
Thus, a pure photon has no charge and mass, even if there are
points inside where $\rho$ and $\rho_m$ can be different from
zero. In addition, a photon proceeds straightly at the speed of
light without dissipation and does not affect the environment
(except for the region directly touched). The curvature of the
space-time is altered at its passage, but not in a significant way
to influence its motion (according to (\ref{eq:gtens}) the
structure of the geodesics is only modified in the direction
orthogonal to that of propagation). There is however a latent
possibility to generating massive charges, and this happens when
the wave is trapped in isolated regions of space. We analyze this
aspect in the coming sections.
\par\smallskip

We finally note that the pressure is a potential. Therefore, our
scalar $p$, after dimensional adjustment, could be put in relation
with a gravitational type potential (see also (\ref{eq:press})),
which is zero for a free single photon and definitely different
from zero when more waves interact. Thanks to (\ref{eq:press}),
$p$ is invariant with respect to coordinate changes. In
\cite{funarol} it is observed that, in wave interactions, there is
a global electromagnetic energy reduction (with respect to the sum
of the free constituents), which is compensated by the creation of
the potential energy associated to $p$. For rotating photons in a
bounded region of space, generating a gravitational environment,
one may impose $\nabla p=0$ at the border (no forces outside the
object). In the coming sections, with the help of explicit
computations, we check exactly what happens in this situation. In
the end, our approach will turn out to be a reasonable proposal
for unifying electromagnetic, mechanical and gravitational
phenomena (we partly agree with the viewpoints expressed in
\cite{chernitskii2}, although our theory is only based on
symmetric tensors).
\par\smallskip

\section{Rotating waves}

We now examine solutions to our model that do not belong to the
subspace of free-waves. The computations are in general very
complicated. Anyway, there are some relatively simple
configurations where explicit solutions are available (see also
\cite{funarol}, chapter 5).
\par\smallskip

In this section, the variables are expressed in cylindrical
coordinates $(r, z, \phi)$ and the solutions will not depend on
$z$. Note that this reference frame is left-handed. We first
define the potentials. For an integer $k\geq 2$ and arbitrary
constants $\omega >0$, $\gamma_0$ and $\gamma_1$, we set:
$${\bf A}= - \Big(\frac{\gamma_1}{\omega} J_{k+1}(\omega r)\sin (c\omega t-k\phi), ~~0~,
~\frac{\gamma_0 r^3}{4k}+ \frac{\gamma_1}{\omega} J_{k+1}(\omega
r)\cos (c\omega t-k\phi)\Big)$$
\begin{equation}\label{eq:poter}
\Phi ~=~-\frac{\gamma_0r^2}{2\omega}- \frac{\gamma_1}{\omega}
J_k(\omega r)\cos (c\omega t-k\phi)
\end{equation}
for $0\leq \phi < 2\pi$, $0\leq r\leq \delta_k /\omega ~$ and any
$z$. In (\ref{eq:poter}), by $J_k$, we denote the $k$-th Bessel
function, that is defined to be a solution to the following
differential equation:
\begin{equation}\label{eq:bes4}
J_k^{\prime\prime}(x)~+~\frac {J_k^\prime (x)} { x}~-~\frac{k^2J_k(x)} {x^2}~+~ J_k(x)~=~0
\end{equation}
The quantity $\delta_k$ turns out to be the first zero of $J_k$,
that, for $k=2$, takes approximately the value (see also table 1):
\begin{equation}\label{eq:del2}
\delta_2~\approx ~5.135622
\end{equation}
The potentials satisfy the Lorenz gauge condition.
From these we get the electromagnetic fields:
$${\bf E}=\left(\frac{\gamma_0r}{\omega}+ \gamma_1
\frac {k J_k(\omega r)} {\omega r}\cos (c\omega t-k\phi)
,~~0,~\gamma_1 J_k^\prime (\omega r)\sin (c\omega
t-k\phi)\right)$$
\begin{equation}\label{eq:cbdiskk}
{\bf B}~=~\frac {1} {c}\Big( 0,~\frac{\gamma_0r^2}{k} +\gamma_1
J_k(\omega r) \cos (c\omega t-k\phi),~~0\Big)
\end{equation}
In order to check the above expressions it is worthwhile to
recall the relations:
$~J^\prime_{k+1}(x)+(k+1)J_{k+1}(x)/x=J_k(x)~$ and
$~J^\prime_k(x)-kJ_k(x)/x=-J_{k+1}(x)$.
\par\smallskip

For $r=\delta_k /\omega $, the time-dependent components of $E_1$
and $B_2$ are zero. Note also that ${\bf E}$ and ${\bf B}$ vanish
for $r=0$, since $J_k(x)$ decays as $x^k$ for $x\rightarrow 0$.
The choice $k=1$ is not permitted because it does not allow to
prolong with continuity the fields up to $r=0$. The idea is to
simulate a $k$-body rotating system in equilibrium. This somehow
explains why the case $k=1$ is not going to produce meaningful
solutions. The wave is constrained in a cylinder $\Sigma$, with
the magnetic field lined up with the axis and the electric field
laying on the orthogonal planes. One has that ${\rm div}{\bf E}$
is constantly equal to $2\gamma_0/\omega$, since the divergence of
the time-dependent part is zero. Moreover, by defining ${\bf
V}=(0, 0, c\omega r/k)$, one finds out that: $D{\bf V}/Dt=
(-(c\omega /k)^2 r, 0 , 0)$.
\par\smallskip

At this point, one can check that all the model equations
(\ref{eq:sfem1}), (\ref{eq:slor1}), (\ref{eq:epres}) are
satisfied, provided we define the pressure as follows:
\begin{equation}\label{eq:pressio}
p=p_0+\frac{c^2\omega
\gamma_0r^2}{k^2}-\frac{\mu\gamma_0^2r^2}{\omega^2}\Big( 1-\frac
{\omega^2r^2}{2k^2}\Big)-\frac{2\mu\gamma_0\gamma_1 r}{k\omega}
J_k^\prime (\omega r)\cos(c\omega t-k\phi )
\end{equation}
where $p_0$ is an arbitrary constant.
This may be checked by substitution in (\ref{eq:epres}) after
recalling that, using  (\ref{eq:bes4}) with $x=\omega r$, the gradient of $p$ is:
$$\frac{\partial p}{\partial r}=\frac{2c^2\omega
\gamma_0r}{k^2}-\frac{2\mu\gamma_0^2r}{\omega^2}\Big( 1-\frac
{\omega^2r^2}{k^2}\Big)-\frac{2\mu\gamma_0\gamma_1 }{kr}
\Big(\frac{k^2}{\omega^2}-r^2\Big)
J_k (\omega r)\cos(c\omega t-k\phi )
$$
\begin{equation}\label{eq:pressiog}
\frac{\partial p}{\partial \phi}~=~-\frac{2\mu\gamma_0\gamma_1 r}{\omega}
J_k^\prime (\omega r)\sin(c\omega t-k\phi )
\end{equation}
Since we have $p\not =0$, we are not in presence of a free-wave.
Due to (\ref{eq:press}), we expect that in the new metric space
defined through (\ref{eq:eins}),  the scalar curvature $R$ is
going to be different from zero, testifying that such a space is
effectively curved. Indeed, such rotating fronts and bent light
rays must be consequences of a space-time deformation, where the
4-velocity $V_\alpha$ should be compatible with the generalized
eikonal equation (\ref{eq:eiko}). Unfortunately, we do not have
the metric tensor in this complicated situation. In order to get
it, one should first create from $(-\Phi ,{\bf A})$ the tensor
$F_{\alpha\beta}$. Successively, one has to build
$T_{\alpha\beta}$ as in (\ref{eq:tens}), (\ref{eq:elec}),
(\ref{eq:mass}), where $g_{\alpha\beta}$ is a generic metric
tensor, and finally solve (\ref{eq:eins}) to find the unknown
$g_{\alpha\beta}$. Note that the pressure in (\ref{eq:pressio}) is
equal to $p_0$ for $r=0$, and that the stationary components of
$({\bf E},{\bf B},{\bf V})$ form a left-handed triplet. It has to
be noticed that the momentum of our rotating wave is preserved
independently of $\omega$: high frequencies ($\omega$ large) are
associated with small diameters $\delta =\delta_k /\omega$, and
vice versa.
\par\smallskip

Although we do not have a rigorous proof, the main idea is
summarized as follows. The $k$ photons rotate around an axis, so
that producing, via Einstein's equation, a gravitational setting
having potential proportional to $p$ (or the scalar curvature $R$,
according to (\ref{eq:press})). On the other hand, the metric
space where such a wave is naturally embedded is such that the
light rays follow circular orbits (geodesics) and the wave-fronts
satisfy an eikonal type equation (see (\ref{eq:eiko})). The
setting may help to better understand the physics of relativistic
rotating bodies (see for instance \cite{rizzi}). By breaking the
equilibrium, we should get $k$ independent free-photons travelling
along straight-lines in an almost flat space with $p=0$ (see
section 4). In this passage, Einstein's equation guarantees global
energy and momentum preservation (see also section 6).
\par\smallskip

\begin{center}
\begin{figure}[!h]\vspace{-.5cm}
\centerline{\includegraphics[width=10.cm,height=5.cm]{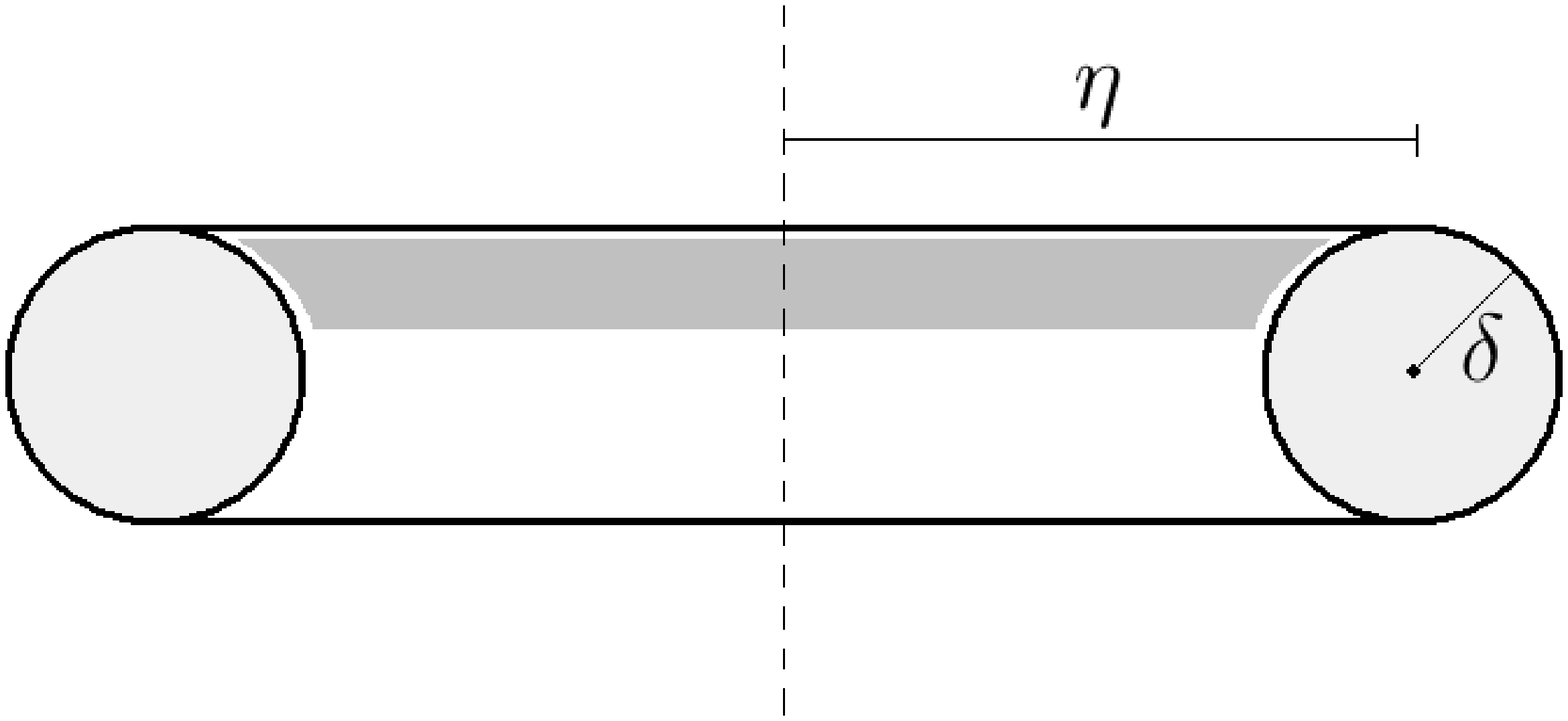}}
\vspace{-.6cm}\centerline{\hspace{4.1cm}\includegraphics[width=8.cm,height=5.cm]{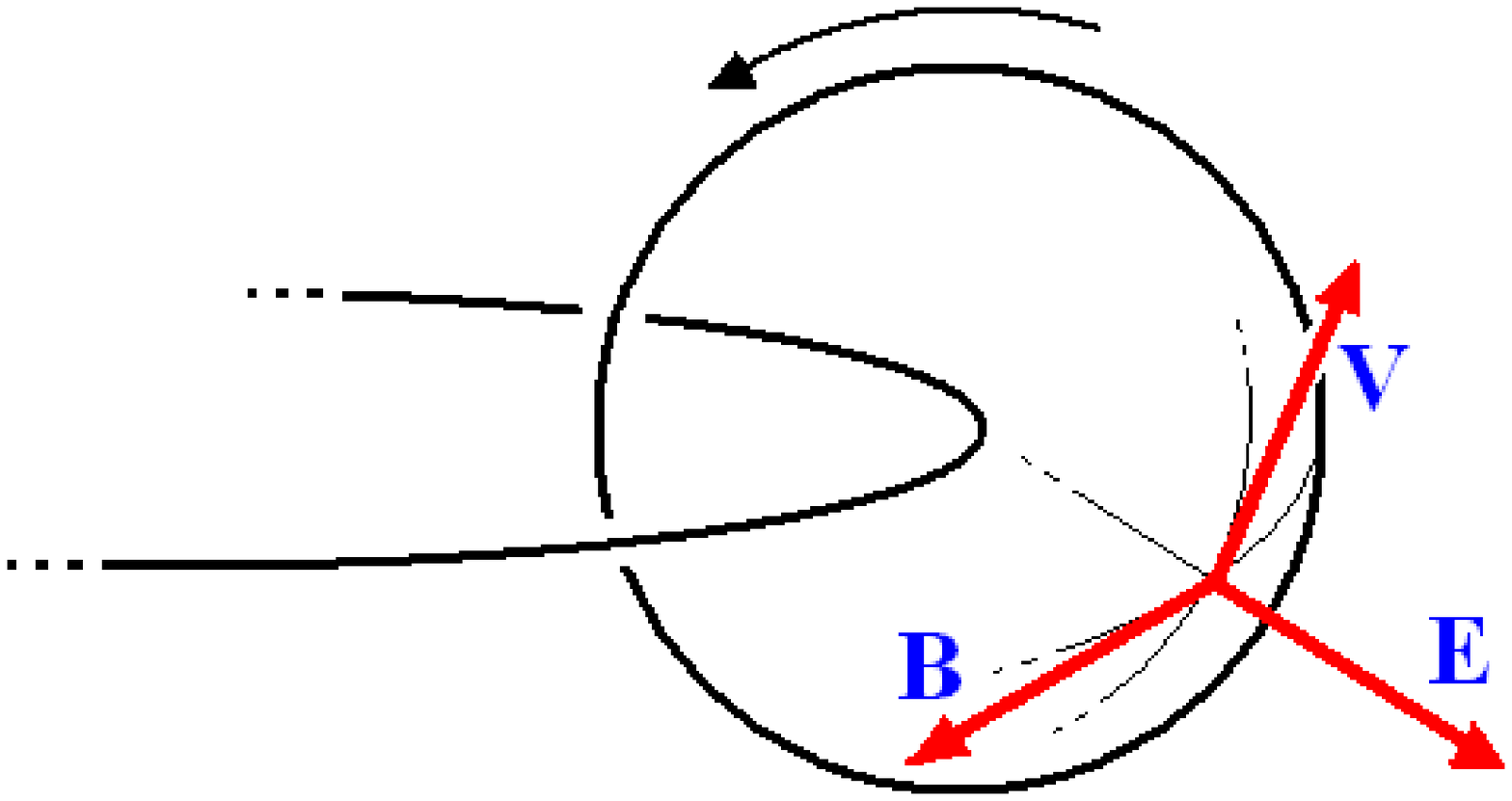}}
\begin{caption}{\small Displacement of the left-handed triplet
$({\bf E}, {\bf B}, {\bf V})$ in the case of a vortex ring.
Stationary electromagnetic fields are overlapped to oscillating
ones, resulting in a rotation around the major circumference. The
first ones provide for charge and magnetic momentum, the second
ones are responsible for the curvature of the space and the
stability of the configuration.}
\end{caption}
\end{figure}
\end{center}

The cylinder $\Sigma$ is unbounded, so that the energy of our wave
turns out to be infinite. We can however build more physical
solutions, by imposing that $\Sigma$ is a toroid with major
diameter $\eta$ and minor diameter $\delta$ (see figure 1). The
stability of vortex rings, from the fluid dynamical point of view,
is impressive. We expect the same to happen to our solutions,
although, due to the difficulty of the problem, such an analysis
looks very hard. The numerical results in \cite{chinosi}, show
that it is possible to build electromagnetic waves constrained in
toroid structures, ranging from the case of a classical ring to a
spherical Hill's type vortex, where the inner hole of the ring is
reduced to a segment. The investigation could help us to
understand some electric phenomena, such as ball lightning or
other self-organized plasmas (see for instance \cite{chen}). For
the usual ring, the study in \cite{chinosi} indicates that, for
$\eta$ slightly bigger than $\delta$, the section $\Omega$ of the
toroid is practically a circle. Thus, one may consider solutions
where the electric field belongs to the vertical section of the
ring, and the magnetic field circulates around the axis (see
figure 1). Hence, in good approximation, the expression of the
fields can be taken equal to that just obtained for the case of
the cylinder. With this setting, we carry out some computations in
the coming section, with the aim to illustrate quantitatively that
electrons could be modelled by electromagnetic vortex rings of the
right size, charge and mass.
\par\smallskip

\section{A model for the electron}

The idea that matter at atomic level has a structure recalling
that of vortex rings dates back to \cite{thomson}. At that time,
however, the knowledge of atomic physics was very poor. Based on
the Maxwell's model, a first effort to describe a spherical
electron with the help of electromagnetic fields was developed in
\cite{poincare} (see also \cite{levi}). The project could not be
realized due to the scarce information available, both theoretical
and experimental. For instance, the concept of {\sl spin} was not
yet introduced.  Moreover, general relativity was developed only
ten years later, and this, to our opinion, must be a fundamental
ingredient for the construction and the stability of the model.
Successively, a purely geometrical approach was considered in
\cite{wheeler}, with the introduction of the so called {\sl
geons}, that is toroid structures in a suitable space-time,
roughly approximated by black holes through the Kerr-Newman metric
(for more recent results, see \cite{arcos}). The literature has
many other scattered contributions on the subject. In the end, all
the proposed models are a bit rough and do not take into account
all the features of a particle. Nevertheless, since now we exactly
know what a photon is, additional elements are available for a
deeper investigation, and the solution we are going to show will
be realistic in all the aspects.
\par\smallskip

Let us consider figure 1. The volume of $\Sigma$ is $2\pi^2\eta\delta^2$, therefore, the
total charge is given by the integral of $\rho ={\rm div}{\bf E}$
over $\Sigma$:
\begin{equation}\label{eq:charge}
q~=~\epsilon_0\int_\Sigma\frac {2\gamma_0}{\omega}~=~\frac {4\pi^2\epsilon_0\eta\delta^2}{\omega}\gamma_0
\end{equation}
Thus, we deduce:
\begin{equation}\label{eq:gamma0}
\gamma_0~=~\frac{q\omega}{4\pi^2\epsilon_0\eta\delta^2}=~\frac{q\omega^3}{4\pi^2\epsilon_0\eta\delta_k^2}
\end{equation}
where $\delta =\delta_k/\omega$, $\delta _k$ being the first zero
of $J_k$. Successively, by integrating the pressure on $\Sigma$,
we obtain:
\begin{equation}\label{eq:ipres}
\int_\Sigma p~=~2\pi\eta \int_\Omega \left[ p_0+\frac{c^2\omega
\gamma_0r^2}{k^2}-\frac{\mu\gamma_0^2r^2}{\omega^2}\Big( 1-\frac
{\omega^2r^2}{2k^2}\Big)\right] r~dr d\phi
\end{equation}
where $\Omega$ is the circle of radius $\delta =\delta_k/\omega$,
and we used the stationary part of the expression in
(\ref{eq:pressio}) (note that the dynamical part has zero average
in $\Omega$). Proceeding with the computation we get:
$$
\int_\Sigma p~=~4\pi^2\eta \left[ \frac{p_0r^2}{2}+
\frac{c^2\omega\gamma_0r^4}{4k^2} ~-~\frac{\mu\gamma_0^2
r^4}{4\omega^2}~+~\frac {\mu \gamma_0^2 r^6}{12k^2}
\right]_0^{\delta_k /\omega}
$$
$$
~=~\frac{2\pi^2\eta p_0\delta_k^2}{\omega^2}+
\frac{\pi^2\eta\gamma_0\delta^4_k}{\omega^3}\left(\frac{c^2}{k^2}~-~
\frac{\mu\gamma_0}{\omega^3}~+~\frac{\mu\gamma_0 A_k}{3\omega^3}
\right)
$$
\begin{equation}\label{eq:ipres2}
~=~\frac{2\pi^2\eta p_0\delta_k^2}{\omega^2}+
\frac{c^2 q A_k}{4\epsilon_0}~+~\frac{\mu
q^2}{16\pi^2 \epsilon_0^2\eta}\left(\frac{A_k}{3}-1\right)
\end{equation}
where we used (\ref{eq:gamma0}) and we set $A_k=\delta_k^2/k^2$.
For convenience, we report in
table 1 the values of $\delta_k$ and $A_k$, for various $k$.
\par\smallskip

\begin{table}[h!]
\label{tab4} \noindent\[
\begin{array}{|c|c|c|c|}
  \hline
\hspace{.1cm}
    ~ k~  &  ~~ ~~~\delta_k ~~~~~ &  ~~ \delta_k/k~~  &  ~ A_k=\delta_k^2/k^2 ~\\
  \hline
  2 & 5.1356 & 2.567  & 6.593 \\
  3 & 6.3802 & 2.126  & 4.522 \\
  4 & 7.5883 & 1.897  & 3.598 \\
  5 & 8.7715 & 1.754  & 3.077 \\
  6 & 9.9361 & 1.656  & 2.742 \\
  7 & 11.086 & 1.583  & 2.508 \\
  8 & 12.225 & 1.528  & 2.335 \\
  9 & 13.354 & 1.483  & 2.201 \\
  10 & 14.476 & 1.447  & 2.095 \\
  11 & 15.590 & 1.417  & 2.008 \\
  12 & 16.698 & 1.391  & 1.936 \\
  13 & 17.801 & 1.369  & 1.875 \\
  14 & 18.900 & 1.350  & 1.822 \\
    \hline
\end{array}
\]
\caption{Values of $\delta_k$ and associated quantities for
various $k$.}
\end{table}

\medskip

After recalling the definition (\ref{eq:made}), we now require
that $\rho_m=0$ for $r=0$ (zero mass density at the center of the
body, in part justified by the fact that $p$ is quadratical with
respect to $r$ near the origin).  This implies that the constant
$p_0$ has to be equal to: $c^2\rho =2\gamma_0c^2/\omega$. We also
remark that a pure rotating wave with $\gamma_0=0$ and
$\gamma_1\not =0$ (see (\ref{eq:cbdiskk})) turns out to have $m
=\int_\Sigma \rho_m =0$ (since $\rho =0$ and $p=0$), according to
the fact that a photon is uncharged and massless, even if it is
constrained in a bounded region of space.
\par\smallskip

Finally, concerning the mass of our toroid, we can write:
\begin{equation}\label{eq:massa}
m~=~\frac{\epsilon_0}{\mu}\int_\Sigma\left(\rho
-\frac{p}{c^2}\right)~=~
-~\frac{q A_k}{4\mu}~-~\frac{q^2}{16\pi^2
\epsilon_0\eta c^2}\left( \frac{A_k}{3}-1\right)
\end{equation}
Let us assume that $2\leq k\leq 5$, so that $A_k>3$ (see table 1).
If $q$ is positive, the above expression is
certainly negative and cannot
represent a mass. If instead $q=-e$ is the electron charge, we
have chances to get something interesting. In such a situation,
the last expression takes the form:
\begin{equation}\label{eq:massa2}
m~=~\frac{A_k}{4}~\frac{e}{\mu}~-~\left( \frac{A_k}{3}-1\right)
\frac{\alpha h}{8\pi^2\eta c}
\end{equation}
where $\alpha =e^2/2hc\epsilon_0$ is the fine structure constant
and $h$ is the Planck constant. It is important to observe that $m$ does
not depend on $\omega$ and $\delta$ (we discuss this aspect later).
\par\smallskip

Successively, we may try to find the major diameter $\eta$. As an
extra condition, we impose that the radial component of $\nabla p$
is zero at the boundary of $\Omega$. In fact, assuming that $p$ is
a kind of gravitational potential (see the end of section 4), this
means that no other external forces are pushing on the boundary
(except for those related to the centripetal acceleration and the
unbalanced Lorentz condition ${\bf E}+{\bf V}\times {\bf B}\not
=0$), producing a sort of equilibrium at the surface of $\Sigma$.
Thus, in order to determine $\eta$, we set to zero the first
relation in (\ref{eq:pressiog}), at the point $\delta_k/\omega$.
Considering that $J_k(\delta_k)=0$, we get the relation:
\begin{equation}\label{eq:calceta}
\frac{c^2}{k^2}~=~\frac{\mu\gamma_0}{\omega^3}( 1-A_k)
\end{equation}
Substituting in (\ref{eq:calceta}) the expression of $\gamma_0$
given in (\ref{eq:gamma0}) and introducing the constant $\alpha$,
one obtains:
\begin{equation}\label{eq:veta}
\eta ~=~\frac{\mu}{e} \frac{\alpha h}{2\pi^2 c}\Big( 1-\frac{1}{A_k}\Big)
\end{equation}
We can go back to (\ref{eq:massa2}). With the new value of $\eta$, we can finally write:
\begin{equation}\label{eq:massa3}
m ~=~\frac{e}{\mu}~ \frac{A_k^2}{6(A_k-1)}
\end{equation}
Let us discuss the case $k=2$. Assuming to know the values of $e$
and $m$, from (\ref{eq:massa3}), we deduce the estimate:
\begin{equation}\label{eq:vum}
\mu~=~1.295~\frac{e}{m}~\approx ~ 2.27\times 10^{11}~{\rm Coulomb/Kg}
\end{equation}
A similar estimate of the constant $\mu$ was recovered in \cite{funarol} by
different arguments. Successively, from (\ref{eq:veta}) we can compute $\eta$:
\begin{equation}\label{eq:vetav}
\eta ~=~\frac{\alpha h A_k}{12\pi^2mc} ~\approx ~.985~\times
10^{-15}~{\rm meters}
\end{equation}
that agrees quite well with the Compton radius of an electron.
\par\smallskip

Let us see what happens to the electromagnetic energy. We take the
stationary parts of the fields in (\ref{eq:cbdiskk}). With
computations similar to those carried out above, we can evaluate
the energy:
$$
{\cal E}~=~\frac{\epsilon_0}{2}\int_\Sigma (\vert{\bf E}\vert^2+c^2\vert{\bf B}\vert^2)
$$
\begin{equation}\label{eq:een}
=~\pi\epsilon_0\eta\int_\Omega\left[\Big(\frac{\gamma_0r}{\omega}\Big)^{\hspace{-.05cm} 2}
+\Big(\frac{\gamma_0 r^2}{k}\Big)^{\hspace{-.05cm} 2}\right]
~=~\Big( 1+\frac{2A_k}{3}\Big)~\frac{\alpha hc}{16\pi^2\eta}
\end{equation}
For $k=2$, using the values already computed, we arrive at:
\begin{equation}\label{eq:eenc}
{\cal E}~=~ \frac{1}{2}\Big( 1+\frac{3}{2A_k}\Big) mc^2~\approx~
.614~ mc^2
\end{equation}
The remaining part of the energy may be provided by the dynamical
part. At this point, one computes $\gamma_1$ in
(\ref{eq:cbdiskk}), in order to get a total energy corresponding
to $~mc^2 =(\epsilon_0 /\mu )\int_\Sigma (\rho c^2 -p)$, that,
written in this form, takes the meaning of a potential energy
(recall that $\rho$ is constant in $\Sigma$).

% If we impose ${\cal E}$ to be
%equal to $mc^2$, where $m$ is the electron mass (in practice, as
%far as the stationary part is concerned, we are requiring that
%$\int_\Sigma T_{00}=0$), one can recover $\eta$:
%\begin{equation}\label{eq:eta}
%\eta~=~\frac{\alpha hK_3}{16\pi^2 mc}~\approx ~ 1.49\times 10^{-15}~{\rm meters}
%\end{equation}
%This agrees with the actual size of an elementary particle.
%Successively, by substituting in (\ref{eq:massa2}), one recovers $\mu$:
%\begin{equation}\label{eq:mu}
%\mu~=~K_4~\frac{e}{m}~\approx ~ 1.25\times 10^{11}~{\rm Coulomb/Kg}
%\end{equation}
%where $~K_4=K_1/(1-K_2/K_3)\approx 0.832$.

\par\smallskip

By analyzing $g_{00}$ in (\ref{eq:rnme}), we can get further
information. For example, if the radius $\eta$ of the particle
coincides with the horizon, so that $(Q/\eta)^2\approx 1$, one
should have: $\chi\approx 32(\pi c^2\eta\epsilon_0 /\mu
e)^2\approx 0.145$.
\par\smallskip

In the end, we were able to prove the following facts.  It is
possible to build electromagnetic vortex rings having the size
($2\eta$), the charge ($-e$) and the mass ($m$) of an electron
(see also \cite{chernitskii}). Note that some computations could
have been made in a different way. For instance, in the definition
of $\rho_m$ (see (\ref{eq:made})), we can include a further
multiplicative adimensional constant. The aim was however to show
that there exists at least an admissible setting.
\par\smallskip

Since our particles are made of rotating photons, it is natural to
associate an intrinsic spin to them, even if they are not
mechanically rotating around an axis. There is also a magnetic
field inside the toroid, having zero dipole moment. This implies
that the particle is `neutral' from the magnetic viewpoint, but,
under external solicitations, it may exhibit a magnetic momentum.
The triplet $({\bf E},{\bf B},{\bf V})$ is left-handed and this
has something to do with the concept of {\sl chirality}. In fact,
the specular image of our particle, obtained by switching the sign
of the charge, produces an anti-particle (the positron), sharing
the same properties of the electron, as it can be proven with
similar computations (compare (\ref{eq:ebpote}) and
(\ref{eq:ebpotea})). When an electron and a positron collide, they
destroy their stable geometrical environment, producing pure
photons departing from the impact site according to standard
mechanical rules.
\par\smallskip

There is no a single solution, but a multitude of them, depending
on the parameter $\omega$. As a matter of fact, maintaining the
amplitude of $\eta$, we can construct thin rings $\Sigma$ having
the sections $\Omega$ rotating at very fast speed. Their charge
and mass densities are high and the volume small. Conversely, one
can build fat rings with lower angular velocity. Probably, there
exists an intermediate solution optimizing the energy, but, with
an analysis at the first order, we were not able to detect it.
Thus, we can associate to our particle infinite frequencies.
However, the whole object has a fixed global diameter $2\eta$,
that can be formally taken as the wave-length of a suitable
fictitious internal clock (see for instance \cite{gouanere}).
These facts are not disturbing however. They make the electron a
very adaptable particle when interacting with the outside world,
explaining why it can emit photons with a continuous spectrum and
escape to finer measurements. In order to justify these
assertions, one has to assume that the bare particle is `dressed
up' with a cloud of photons, bringing different frequencies and
responsible for the transmission of the electric charge to the
environment. At this point, the discussion becomes more delicate
and the interested reader is addressed to \cite{funarol}, chapter
6, and \cite{funaroa}, where the author gives his personal
interpretation on issues regarding the quantization of matter.
\par\smallskip

Switching the sign of both the electric and magnetic fields,
should bring to a positive left-handed particle. Its structure is
going to be quantitatively different from that of the electron,
since the forces in (\ref{eq:epres}) find their balance in an
alternative way. We do not think however that the standard ring
will be stable in this case, while we guess that a Hill's type
vortex would be more appropriate (see \cite{chinosi}). On the
other hand, if also protons can be described by our model, we do
expect they assume complicated shapes (see \cite{miller}) and
display peculiar substructures, as it is revealed by experiments.

\par

\end{document}